\documentclass[11pt,twoside,a4paper,cr,final]{cms-tdr}
\def\svnVersion{64946M}

\begin{document}

\hyphenation{had-ron-i-za-tion}
\hyphenation{cal-or-i-me-ter}
\hyphenation{de-vices}

\RCS$Revision: 62698 $
\RCS$HeadURL: svn+ssh://tdahms@svn.cern.ch/reps/tdr2/papers/XXX-08-000/trunk/XXX-08-000.tex $
\RCS$Id: XXX-08-000.tex 62698 2011-06-21 00:28:58Z alverson $
\newcommand{\ee}{\ensuremath{e^+e^-}\xspace}
\newcommand{\mumu}{\ensuremath{\mu^+\mu^-}\xspace}
\newcommand{\eexp}[1]{\ensuremath{{\rm e}^{#1}}\xspace}
\newcommand{\Jpsi}{\JPsi}

\newcommand{\eq}[1]{~(\ref{#1})\xspace}
\newcommand{\fig}[1]{{figure~\ref{#1}}\xspace}
\newcommand{\tab}[1]{\ref{#1}\xspace}

\renewcommand{\eg}{\emph{e.\,g.}\xspace}%
\renewcommand{\ie}{\emph{i.\,e.}\xspace}%

\newcommand{\pp}{{\ensuremath{pp}}\xspace}
\newcommand{\PbPb}{\ensuremath{\rm{PbPb}}\xspace}
\newcommand{\AuAu}{\ensuremath{\rm{AuAu}}\xspace}

\newcommand{\sqrts}{\ensuremath{\sqrt{s}}\xspace}
\newcommand{\sqrtsnn}{\ensuremath{\sqrt{s_{NN}}}\xspace}

\newcommand{\pT}{\pt}

\newcommand{\pythia}{{\sc Pythia}\xspace}
\newcommand{\hydjet}{{\sc Hydjet}\xspace}
\newcommand{\cascade}{{\sc Cascade}\xspace}

\cmsNoteHeader{2011-111} 
\title{Charmonium production measured in PbPb and pp collisions by CMS}

\address[llr]{Laboratoire Leprince-Ringuet (LLR), \'{E}cole Polytechnique, 91128 Palaiseau, France}
\author[llr]{Torsten Dahms\footnote{The author presenting these results
    received funding from the European Research Council under the FP7
    Grant Agreement no. 259612.}, for the CMS collaboration}

\date{\today}

\abstract{
  The Compact Muon Solenoid (CMS) is fully equipped to measure hard
  probes in the di-muon decay channel in the high multiplicity
  environment of nucleus-nucleus collisions. Such probes are
  especially relevant for studying the quark-gluon plasma since they
  are produced at early times and propagate through the medium,
  mapping its evolution. CMS is able to distinguish non-prompt from
  prompt \Jpsi in \pp and \PbPb collisions. We report here the nuclear
  modification factor of prompt \Jpsi in \PbPb as a function of
  transverse momentum, rapidity, and the number of nucleons
  participating in the collision.
}

\hypersetup{%
pdfauthor={CMS Collaboration},%
pdftitle={Charmonium production measured in PbPb and pp collisions by CMS},%
pdfsubject={CMS},%
pdfkeywords={CMS, physics, heavy-ions, dimuons, quarkonia}}

\maketitle 

The goal of the SPS, RHIC, and LHC heavy-ion programmes is to probe
the existence and study the properties of the quark-gluon plasma
(QGP), a state of deconfined quarks and gluons. In such a state it was
predicted~\cite{matsui-satz} that quarkonium production is suppressed
by Debye screening of the heavy quark binding potential. Therefore,
the observation of such suppression is considered one of the most
promising signatures of the creation of the QGP. At LHC energies and
luminosities, the contribution of non-prompt \Jpsi from B-hadron
decays to the inclusive \Jpsi cross-section is non-negligible and
needs to be considered. CMS has been able to separate prompt from
non-prompt \Jpsi not only in \pp collisions at \sqrts =
7\,TeV~\cite{BPH-10-002}, but also in the much more challenging
environment of \PbPb collisions at \sqrtsnn =
2.76\,TeV~\cite{HIN-10-006}. In this proceedings, the CMS measurement
of prompt \Jpsi production in \PbPb collisions are reported. The
results are presented as nuclear modifications factors $R_{AA}$ based
on a comparison to the yield measured in a \pp reference run at the
same \sqrtsnn.

The central feature of CMS is a superconducting solenoid, of 6\,m
internal diameter, providing a field of 3.8\,T. Within the field volume
are the silicon pixel and strip tracker, the crystal electromagnetic
calorimeter (ECAL) and the brass/scintillator hadron calorimeter
(HCAL). Muons are measured in gas-ionization detectors embedded in the
steel return yoke. In addition to the barrel and endcap detectors, CMS
has extensive forward calorimetry. The muons are measured in the
pseudorapidity window $|\eta|< 2.4$, with detection planes made of
three technologies: Drift Tubes, Cathode Strip Chambers, and Resistive
Plate Chambers. Matching the muons to the tracks measured in the
silicon tracker results in a transverse momentum resolution better
than 1.5\% for \pT smaller than 100\,\GeVc. A much more detailed
description of CMS can be found elsewhere~\cite{JINST}.

\begin{figure}[!htp]
  \centering
  \includegraphics[width=0.4\textwidth]{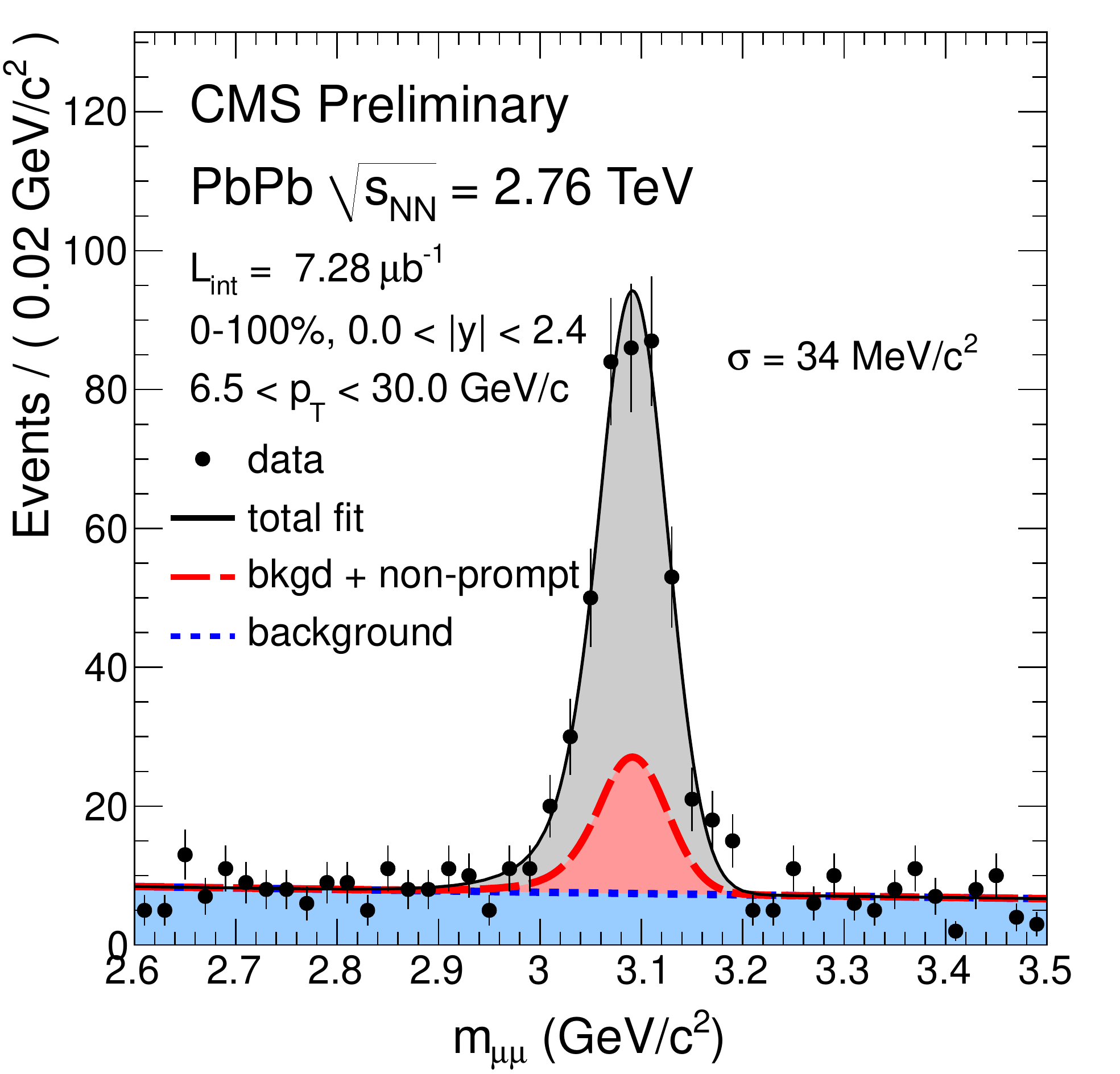}\hspace{1em}
  \includegraphics[width=0.4\textwidth]{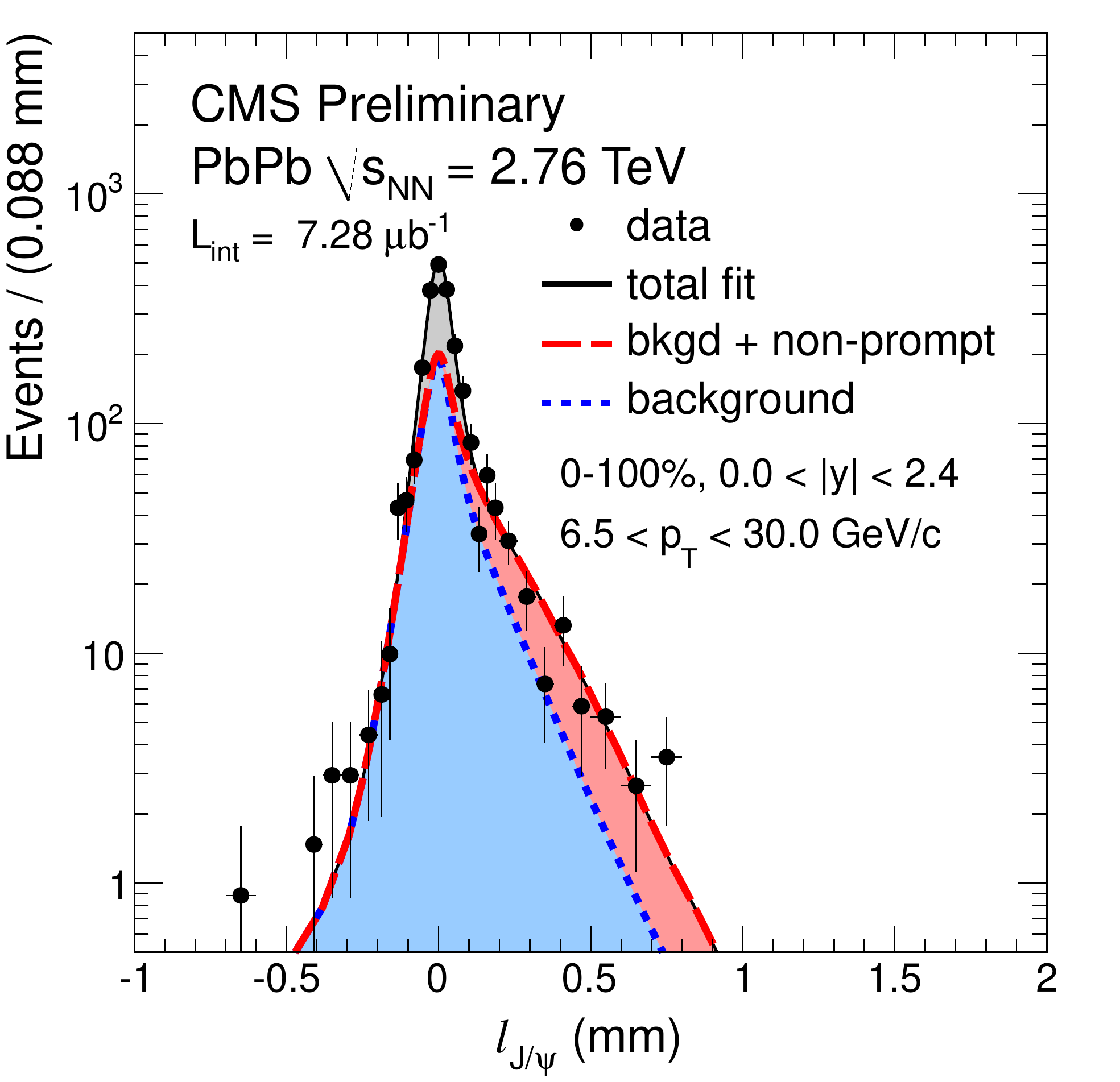}
  \caption{Invariant mass spectrum (left) and pseudo-proper decay
    length (right) of \mumu pairs with $6.5<\pT<30$\,\GeVc and $|y| <
    2.4$ integrated over the centrality range 0--100\%. The data
    (black circles) are overlaid with the projections of the
    2-dimensional fit. The different contributions are background
    (dotted blue line), non-prompt \Jpsi (dashed red line), and the
    sum of background, non-prompt and prompt \Jpsi (solid black
    line).}
  \label{fig:jpsifit}
\end{figure}

The \Jpsi analysis in \PbPb collisions at \sqrtsnn = 2.76\,TeV is
based on a double-muon triggered data set recorded during the LHC
heavy-ion run in 2010 and corresponds to a total integrated luminosity
of ${\cal L}_{\rm int} = 7.28\,\mu{\rm b}^{-1}$. The analysis follows
closely the one performed in \pp collisions at \sqrts = 7 TeV whose
results are reported in~\cite{BPH-10-002}. In the offline analysis,
muon candidates are reconstructed as tracks in the muon
detectors. These tracks are then matched to tracks reconstructed in
the silicon tracker, based on algorithms optimized for the high
multiplicity environment of heavy-ion
collisions~\cite{Chatrchyan:2011sx}. Muons are further identified by
cuts on the quality of the tracks reconstructed in the silicon tracker
as well as the ones in the muon chambers. Opposite-sign muon pairs
with an invariant mass between 2.6\,\GeVcc and 3.5\,\GeVcc and a good
common vertex probability are considered for the \Jpsi analysis. Their
invariant mass distribution is shown in the left panel of
\fig{fig:jpsifit} for \mumu pairs with $6.5<\pT<30$\,\GeVc integrated
over the rapidity range $|y| < 2.4$ and the centrality range
0--100\%. The \Jpsi peak is well reconstructed with a mass resolution
of 34\,\MeVcc. The data are fitted with a Crystal Ball function plus
an exponential for the background.
To distinguish prompt from non-prompt \Jpsi, the pseudo-proper decay
length $\ell_{\Jpsi} = L_{xy} \cdot m_{\Jpsi}/\pT$ is computed for
each \Jpsi candidate. $L_{xy}$ is the most probable transverse decay
length in the laboratory frame defined as:
\begin{equation}
  L_{xy} = \frac{\hat{u}^T\sigma^{-1}\vec{x}}{\hat{u}^T\sigma^{-1}\hat{u}}
\end{equation}
where $\vec{x}$ is the vector joining the dimuon vertex and the
collision vertex of the event, $\hat{u}$ is the unit vector of the
\Jpsi \pT, and $\sigma$ is the sum of the primary and secondary vertex
covariance matrices~\cite{BPH-10-002}. The fraction of non-prompt
\Jpsi is determined by an unbinned maximum likelihood fit in bins of
\pT, rapidity, and centrality~\cite{HIN-10-006,mihee}. The invariant
mass and the pseudo-proper decay length distributions are fitted
simultaneously. The projection along $\ell_{\Jpsi}$ is shown in the
right panel of \fig{fig:jpsifit} in which the data are overlaid with
the different contributions to the fit: prompt signal, non-prompt
signal and background.  In March 2011, a reference sample of \pp
collisions at \sqrts = 2.76\,TeV has been collected. It corresponds to
an integrated luminosity of ${\cal L}_{pp} = 225\,{\rm nb}^{-1}$. The
data are reconstructed with the same heavy-ion optimized
reconstruction algorithms and the analysis is performed identically to
the \PbPb analysis.

Based on the \Jpsi raw yields $N_{\PbPb}(\Jpsi)$ and $N_{\pp}(\Jpsi)$
measured in \PbPb and \pp collisions, respectively, the nuclear
modification factor
\begin{equation}
  R_{AA} = \frac{{\cal L}_{pp}}{T_{AA} N_{\rm MB}}\frac{N_{\PbPb}(\Jpsi)}{N_{\pp}(\Jpsi)}\frac{\varepsilon_{\pp}}{\varepsilon_{\PbPb}}
\end{equation}
has been calculated. The only correction applied is a multiplicity
dependent difference of the trigger and reconstruction efficiencies in
\PbPb and \pp collisions
($\frac{\varepsilon_{\pp}}{\varepsilon_{\PbPb}}=1.17$ for the most
central bin). $T_{AA}$ is the nuclear overlap function and $N_{MB}$
the number of sampled minimum bias \PbPb collisions. The trigger and
reconstruction efficiencies are calculated using Monte Carlo
simulations where a \pythia signal event is embedded into a background
event simulated with \hydjet~\cite{hydjet}. These events are processed
through the full CMS trigger emulation and event reconstruction
chain. The results are cross checked with real data using a \emph{tag
  and probe} technique~\cite{HIN-10-006}.

\begin{figure}[!htp]
  \centering
  \includegraphics[width=0.4\textwidth]{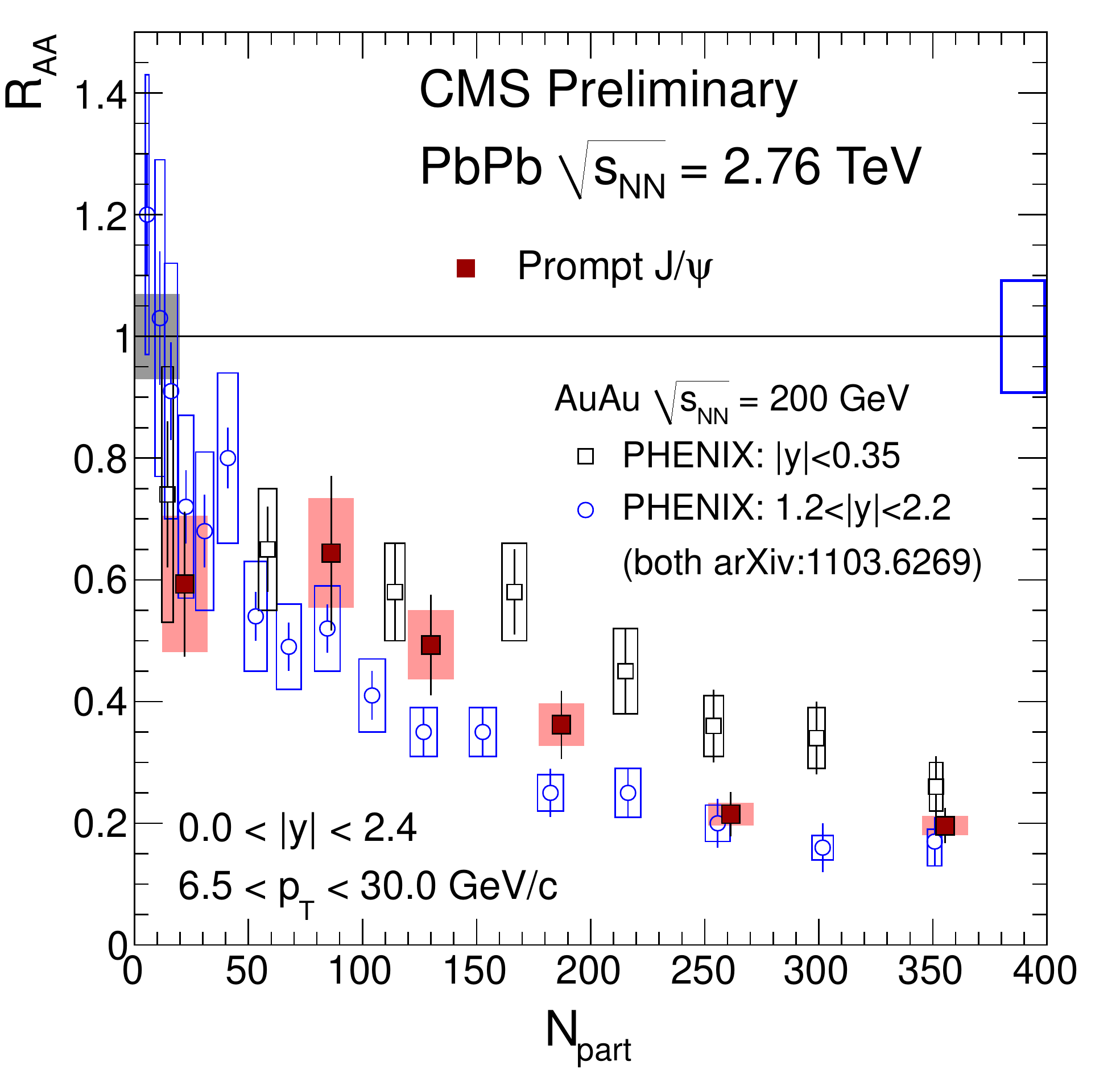}\hspace{1em}
  \includegraphics[width=0.4\textwidth]{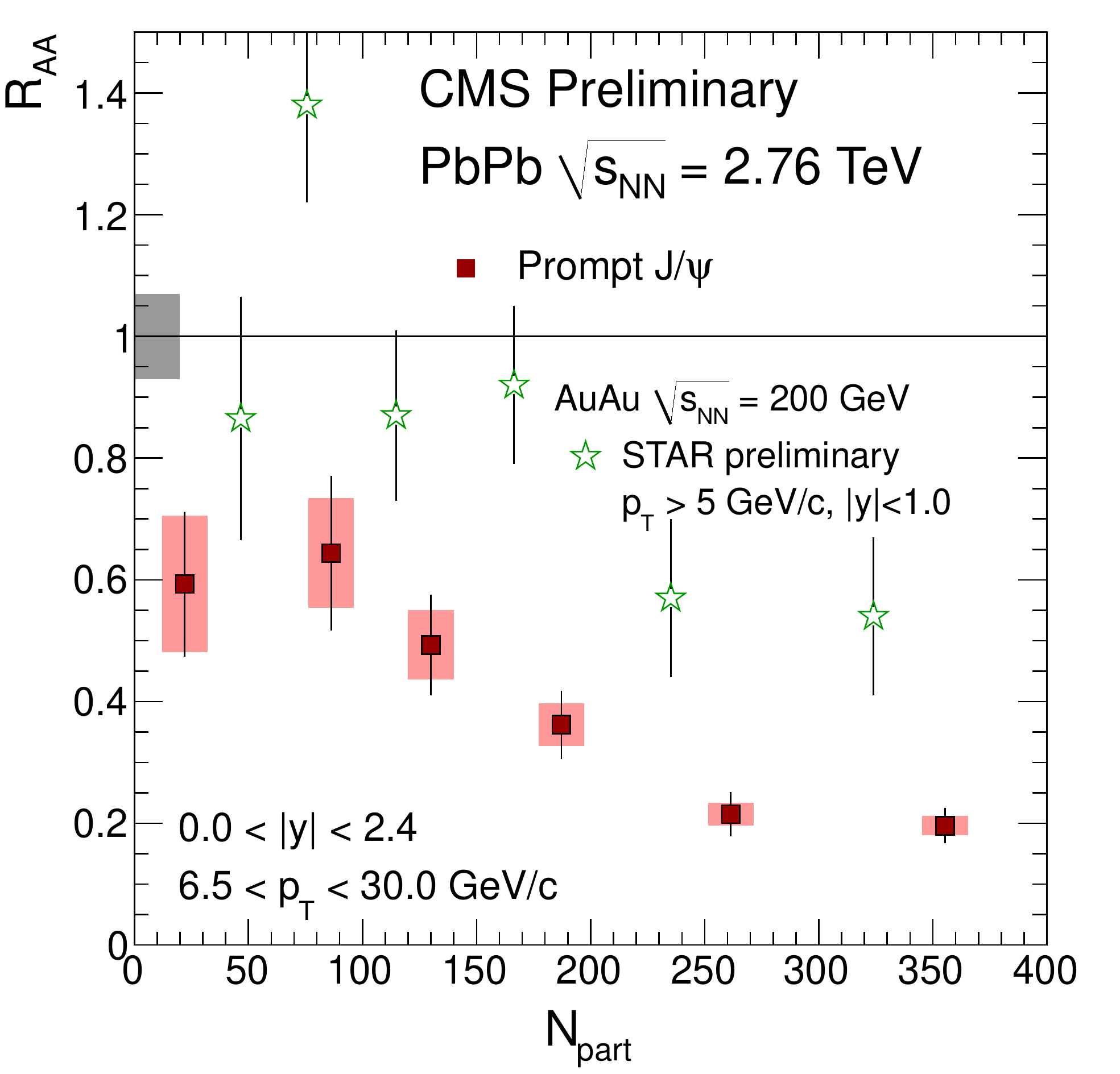}
  \caption{$R_{AA}$ of prompt \Jpsi (red squares) as function of
    $N_{\rm part}$. The left panel shows the comparison to PHENIX data
    at mid- (open black squares) and at forward rapidity (open blue
    circles). The right panel compares to STAR data (green stars).}
  \label{fig:raaNpart}
\end{figure}

\begin{figure}[!htp]
  \centering
  \includegraphics[width=0.4\textwidth]{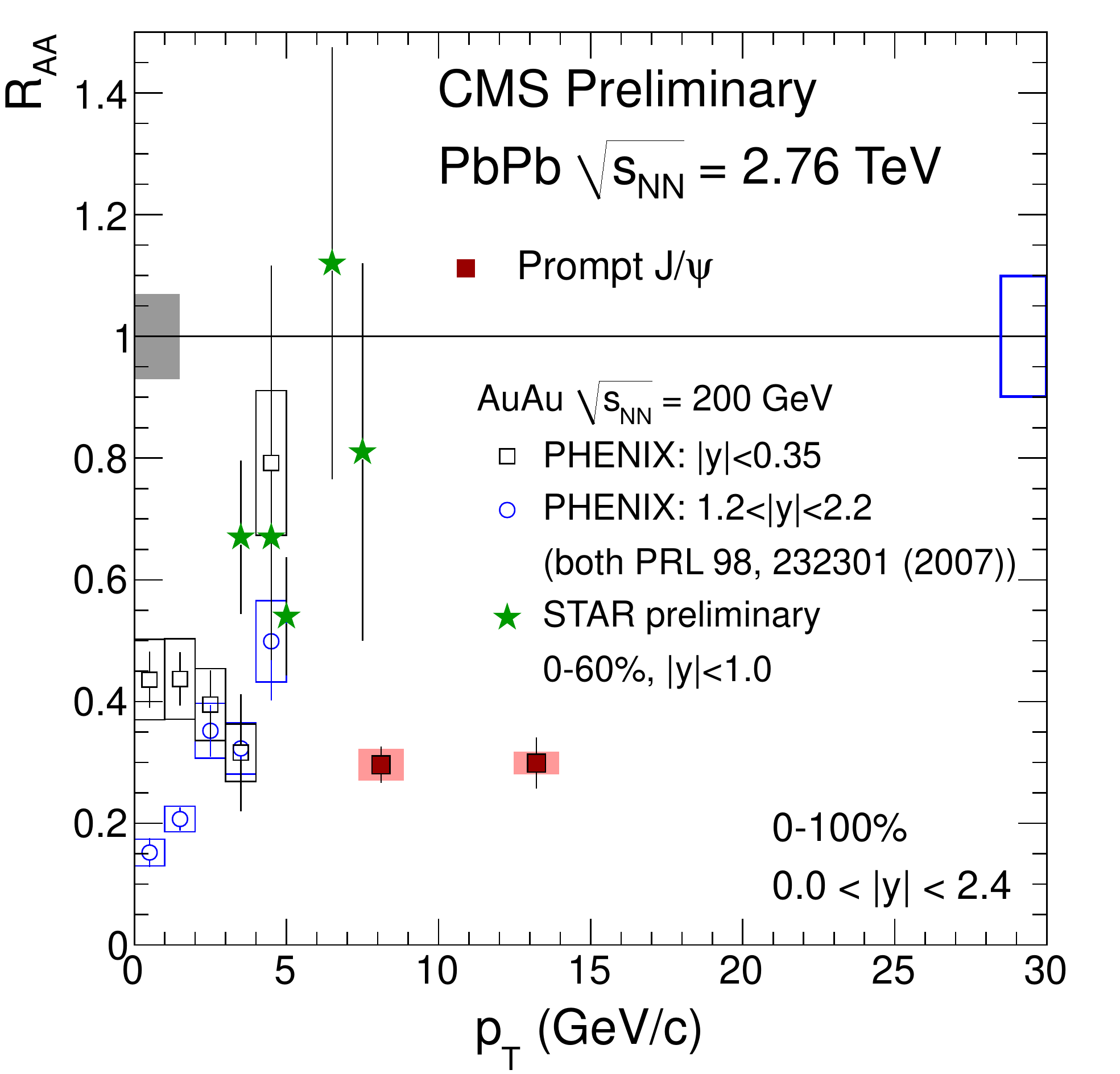}\hspace{1em}
  \includegraphics[width=0.4\textwidth]{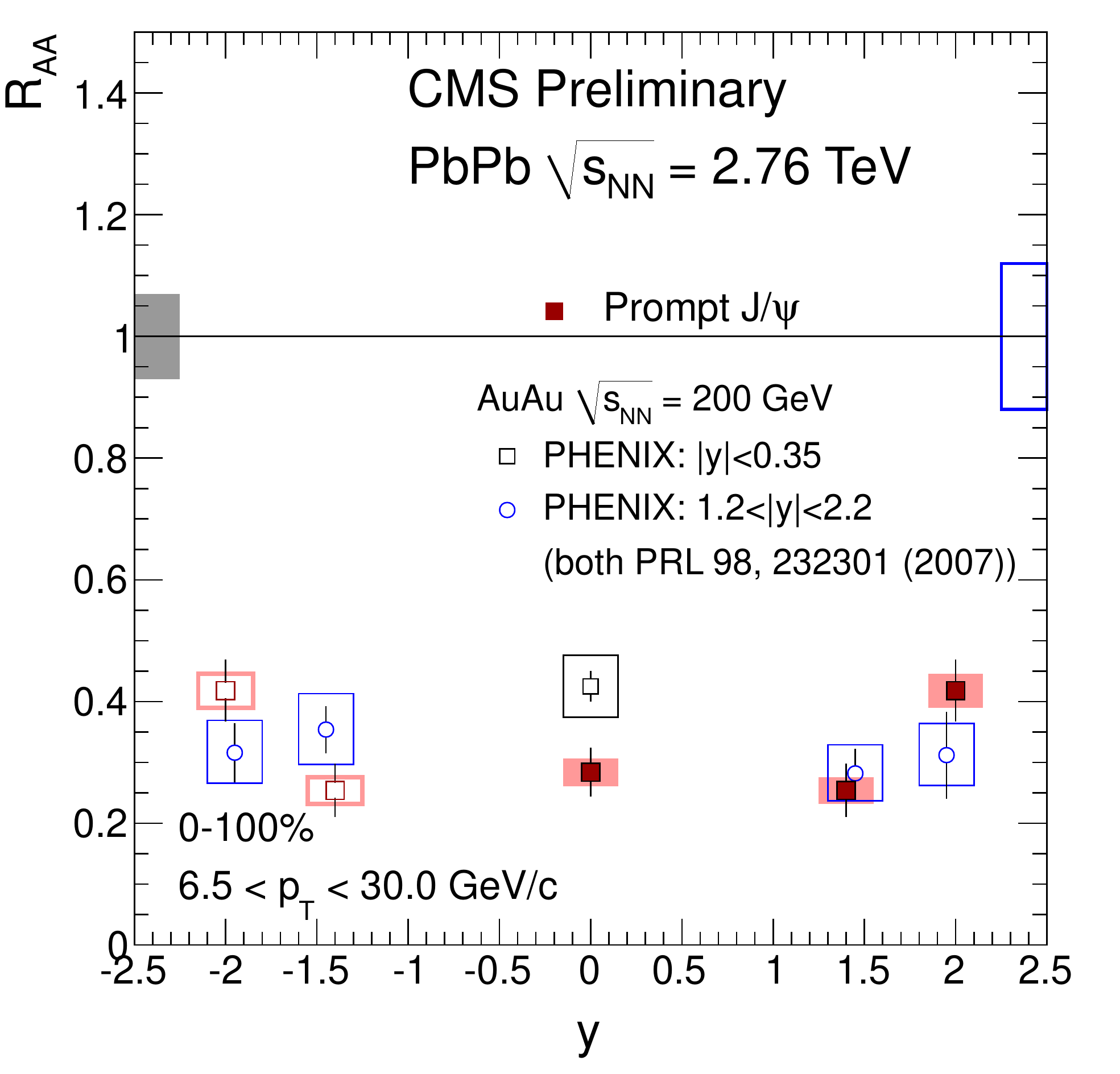}
  \caption{$R_{AA}$ of prompt \Jpsi (red squares) as function of \pT
    (left) and rapidity (right). The data are compared to PHENIX and
    STAR results.}
  \label{fig:raaPtY}
\end{figure}

The resulting $R_{AA}$ of prompt \Jpsi with $6.5<\pT<30$\,\GeVc and
$|y|<2.4$ is shown in \fig{fig:raaNpart} as function of the number of
participating nucleons ($N_{\rm part}$). A centrality dependent
suppression of prompt \Jpsi production is observed. In the 10\% most
central events, the suppression reaches a factor of 5 with respect to
the \pp reference. The suppression decreases towards peripheral events
to a factor 1.6 in the 50--100\% centrality bin. The results are
compared to measurements in \AuAu collisions at \sqrtsnn = 200 GeV by
PHENIX~\cite{phenix} and STAR~\cite{star}. The PHENIX results,
measured in two rapidity ranges ($|y|<0.35$ and $1.2<|y|<2.2$) at much
smaller \pT, show a surprisingly similar suppression pattern. The STAR
result of the \Jpsi $R_{AA}$ for $\pT > 5$\,\GeVc, however, shows less
suppression than measured by CMS.
The \pT and rapidity dependence of the $R_{AA}$ are shown in
\fig{fig:raaPtY}. No strong \pT dependence is observed in the measured
range. There is some indication for a decrease of the suppression at
forward rapidity. As for the centrality dependence, the data are
compared to results from PHENIX and STAR. The rapidity dependence of
the suppression measured by PHENIX at low \pT seems to indicate the
opposite trend with respect to the CMS measurement at high \pT. The
STAR data show less suppression of high \pT \Jpsi at RHIC energies
than at the LHC.

CMS has separated the prompt \Jpsi from the non-prompt contribution
due to B-hadron decays. The nuclear modification factor of prompt
\Jpsi has been measured as function of centrality, \pT, and rapidity
in PbPb collisions at \sqrtsnn = 2.76\,TeV. A suppression of prompt
\Jpsi has been observed, which increases with centrality up to a
factor of 5. CMS has also measured the $R_{AA}$ of non-prompt \Jpsi
which gives access to the in-medium energy loss of
b-quarks~\cite{HIN-10-006,mihee}.

\end{document}